\documentclass[aps,twocolumn,superscriptaddress,prl,longbibliography]{revtex4-2}

\usepackage{graphicx}
\usepackage{placeins} 
\usepackage{amssymb,amsmath,amsfonts,bm}
\usepackage{subcaption}
\usepackage[normalem]{ulem}
\usepackage{hyperref}
\usepackage[dvipsnames]{xcolor}
\hypersetup{urlcolor=RoyalBlue, colorlinks=true,citecolor=RoyalBlue,linkcolor=RoyalBlue} 
\usepackage{physics}
\usepackage[final]{showkeys}
\usepackage{float}

\begin{document}
\title{$S=1$ pyrochlore magnets with competing anisotropies: A tale of two Coulomb phases, $Z_2$ flux confinement and $XY$-like transitions.}
\author{Jay Pandey}
\affiliation{Tata Institute of Fundamental Research, Mumbai India 400005}
\author{Kedar Damle}
\affiliation{Tata Institute of Fundamental Research, Mumbai India 400005}

\date{\today}

\begin{abstract}
We argue that the low-temperature physics of $S=1$ pyrochlore magnets with a predominantly Ising-like easy-axis exchange coupling $J$ that favors the local tetrahedral body diagonals, and a comparably large easy-plane single-ion anisotropy $\Delta =J + \mu$ ($|\mu| \ll J$) that favors the plane perpendicular to these local axes will exhibit interesting new phenomena due to the competition between $J$ and $\Delta$. In the $T/J \rightarrow 0$ limit, we find three low temperature phases as a function of $\mu/T$: a short-range correlated paramagnetic phase, and two topologically-distinct Coulomb liquids separated by a $Z_2$ flux confinement transition. Both Coulomb liquids are described at long-wavelengths by a fluctuating divergence-free polarization field and have characteristic pinch-point singularities in their structure factor. In one Coulomb phase, the flux of this polarization field is confined to {\em even} integers, while it takes on all integer values in the other Coulomb phase. 
Experimental realizations with $|\mu| \ll J$ and negative are predicted to exhibit signatures of a transition from a flux-deconfined Coulomb phase to the flux-confined Coulomb phase as they are cooled below   $T_{c_2} \approx 1.57|\mu|$, while realizations with positive $\mu \ll J$ will show signatures of a transition from a flux-deconfined Coulomb liquid to a short-range correlated paramagnet via a continuous $XY$-like transition at $T_{c_1} \approx 0.98 \mu$. 
\end{abstract}

\keywords{Suggested keywords}
\maketitle

\section{Introduction}
Insulating magnets in which the dominant magnetic interactions compete due to the geometry of the lattice, are known to exhibit interesting low-temperature phenomena~\cite{Mila_Mendels_Lacroix_2011book}. These include ``spin-liquid'' 
states~\cite{Broholm_Cava_Kivelson_Nocera_Norman_Senthil_Science_2020,
Clark_Abdeldaim_Annual_Review_of_Material_Research_2021,Zhou_Kanoda_Ng_RevModPhys_2017,
Knolle_Moessner_Annual_Review_of_Condensed_Matter_Physics_2019}
whose description is most natural in terms of emergent degrees of freedom rather than the microscopic spins 
themselves. These {\em geometrically-frustrated} magnets provide an arena for interesting new physics in part because this competition between the leading magnetic interactions results in a large near-degeneracy of low-energy configurations~\cite{Anderson_1956}. 
When quantum fluctuations are important, this can give rise to topologically ordered quantum states with entirely new kinds of collective modes as well as particles with fractional quantum numbers~\cite{Wen_RevModPhys_2017,Wen_Science_2019,
Benton_PRb_2012,Banerjee_etal_PRL_2008,Hermele_Fisher_Balents_PRB_2004}. 
Even when thermal fluctuations dominate over quantum effects, the low-temperature behavior in some cases is  best described not in terms of the microscopic spins themselves, but in terms of emergent degrees of freedom and their thermal fluctuations~\cite{Castelnovo2008,Jaubert_Holdsworth_2009,Kadowaki_etal_2009,
Rehn_Sen_Damle_Moessner_PRL_2016,
Rehn_Sen_Moessner_PRL_2017}. 

Classical spin-ice materials such as Ho$_2$Ti$_2$O$_7$, in which the magnetic ions lie on the vertices of a network of corner-sharing tetrahedra forming the pyrochlore lattice,  are arguably the best-studied examples of this physics~\cite{Bramwell_Harris_review2020,Bramwell_Gingras_review_2001_science,Gardner_Gingras_Greedan_2010}. 
\begin{figure*}
    (a)\includegraphics[width=0.3\linewidth]{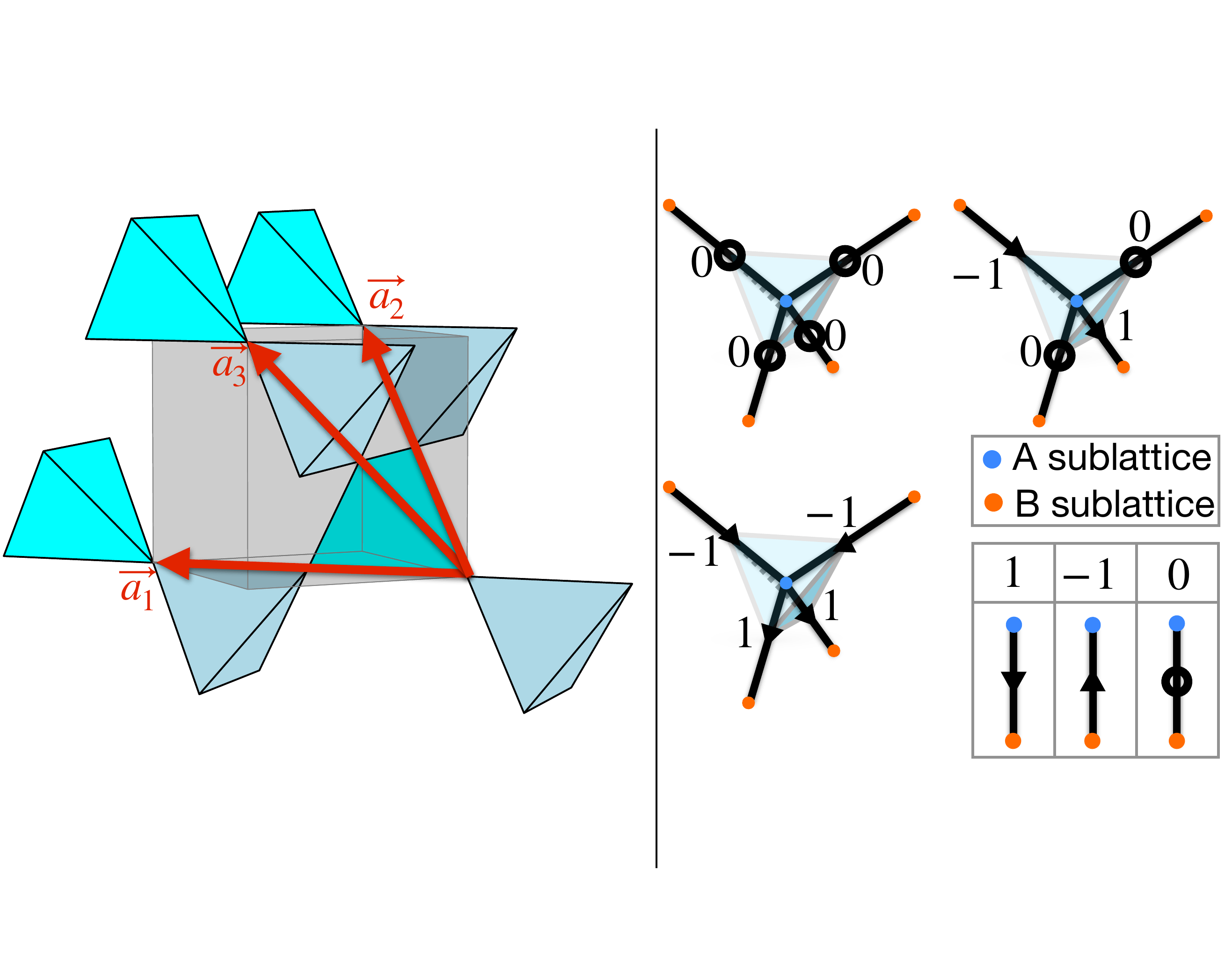}
    (b)\includegraphics[width=0.3\linewidth]{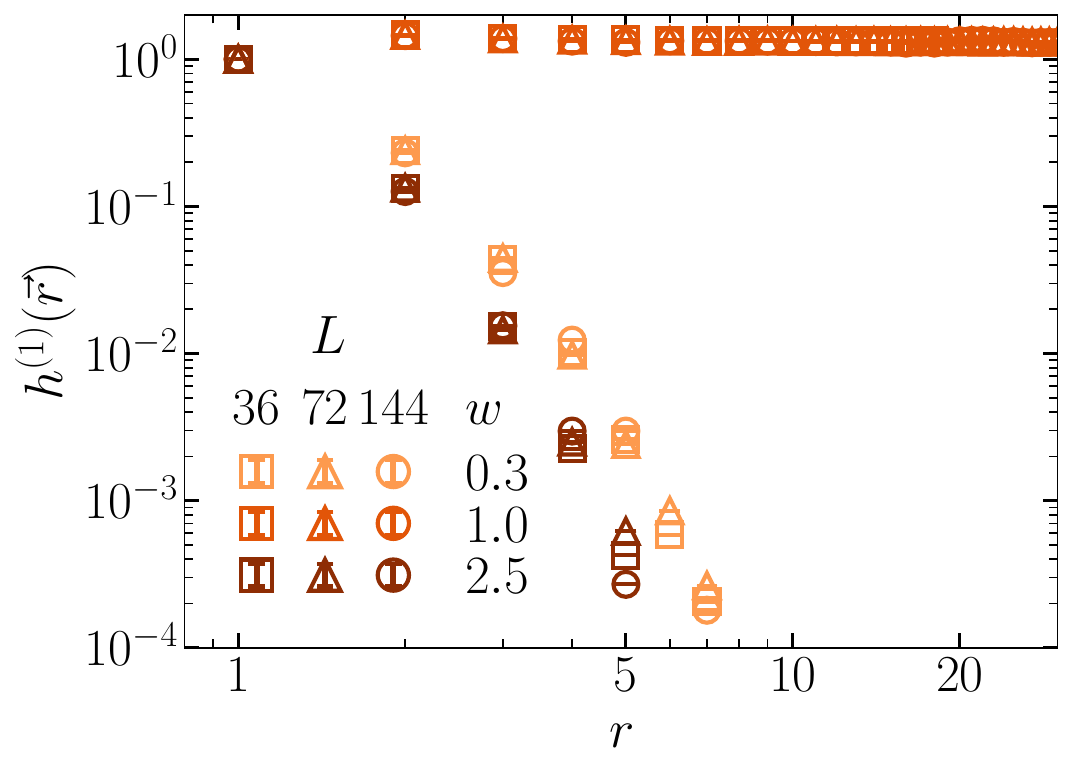}
    (c)\includegraphics[width=0.3\linewidth]{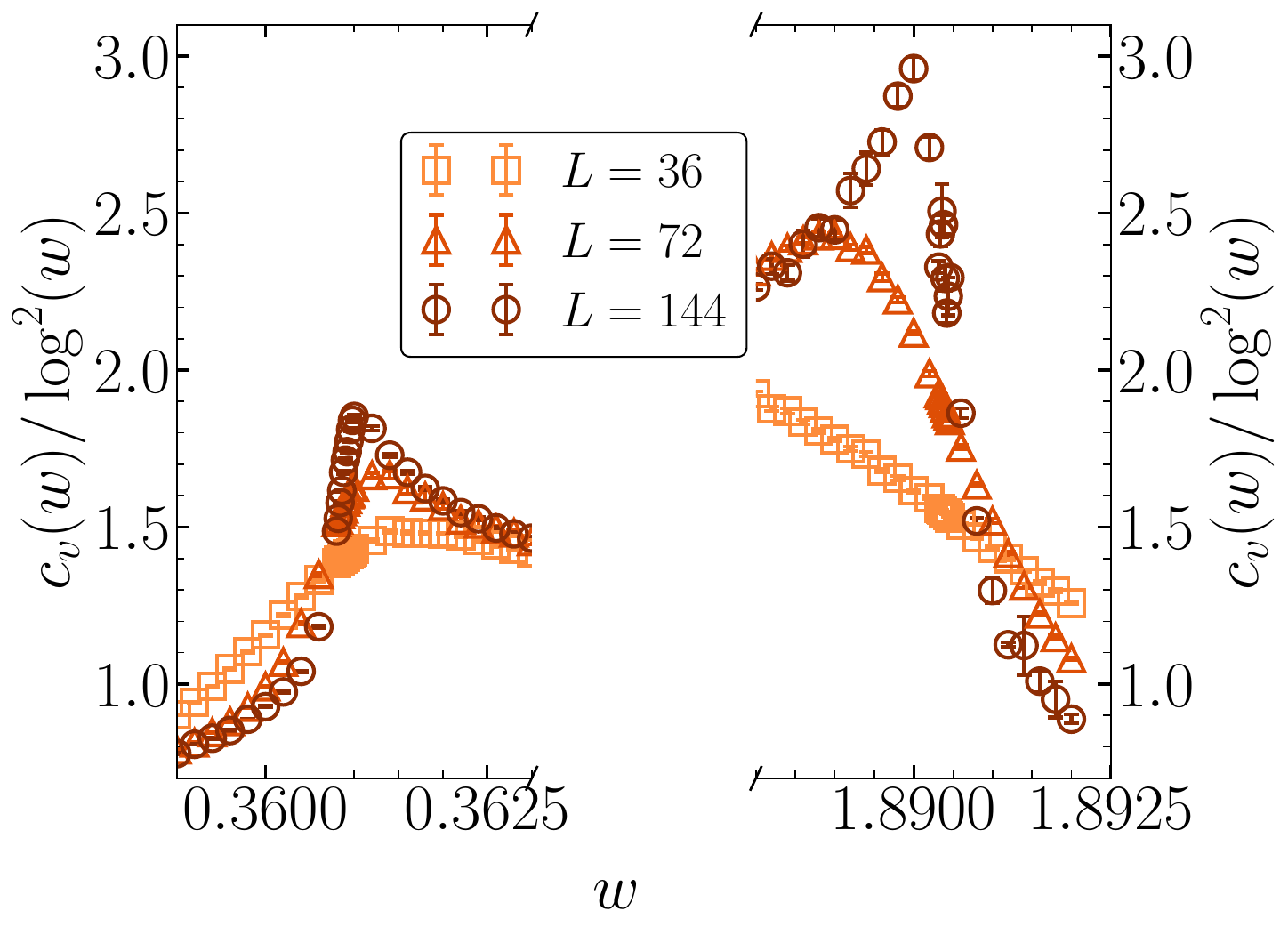}
    \caption{(a) Principal axes (Bravais lattice translations) of the pyrochlore lattice, the low-energy spin configurations of an up-pointing tethrahedron, and the mapping to a divergence-free polarization field (b) The test $\pm 1$ charge correlator, measured by the histogram $h^{(1)}(\vec{r})$ of head-to-tail distances of the unit-charge worm, is deconfined for intermediate $w$ and confined both for $w \gg 1$ and $w \ll 1$, indicating the presence of three distinct phases. (c) The specific heat shows two sharp features, consistent with the presence of two distinct thermodynamic phase transitions. }
    \label{fig:SchematicTestCorrelatorSpecificHeat}
\end{figure*}
There are two key ingredients that lead to classical spin ice behavior in many of these spin-ice materials: First,  at the level of a single magnetic ion, the combined effects of a large spin-orbit coupling and the crystal-field potential results in a high spin (total angular momentum) state described by an effective spin-$S$ degree of freedom $\vec{S}$ that sees a strong easy-axis anisotropy, with the easy-axis ``$z$'' for each site being oriented along the tetrahedral body diagonal that passes through the site~\cite{Rosenkaraz_etal_2000,Harris_etal_1997_PRL}. 
This forces the magnetic exchange interactions and dipolar couplings between these magnetic moments to act predominantly within the low-energy doublet corresponding to $S^z = \pm S$, leading to an essentially classical description in terms of Ising pseudospins with frustrated interactions; quantum fluctuations are negligible since the transverse couplings between these Ising spins are severely suppressed by this single-ion physics. 
Thus, both the strong exchange anisotropy and the resulting classical behavior are in effect {\em consequences} of the single-ion anisotropy in such materials~\cite{Rosenkaraz_etal_2000,Harris_etal_1997_PRL}.
 
Here, we demonstrate that new low-temperature phases and transitions between them can arise in effective spin $S = 1$ pyrochlore magnets if this is not the case, {\em i.e.}, if the dominant effective exchange coupling $J$ mainly couples neighboring $S^z$ (with $z$ defined as above) {\em although} the $S=1$ moments have a comparably strong single-ion {\em easy-plane} anisotropy $\Delta  = J + \mu $ (with $|\mu| \ll J$) that favours the plane perpendicular to the local $z$ axis.
In the $T/J \rightarrow 0$ limit with $\mu/T$ held fixed, we find that this competition leads to three distinct low temperature phases: A short-range correlated paramagnet and two topologically-distinct Coulomb liquids. 

The two Coulomb liquids both admit an effective field theory description in terms of a coarse-grained divergence-free polarization field and have characteristic pinch-point singularities~\cite{Bramwell_etall_2001,Fennel_etal_2004,Fennel_etalScience2009} in their structure factor, but are separated by a $Z_2$ flux confinement transition: In the $Z_2$-confined Coulomb phase, the flux of this polarization field is confined to {\em even} integers, while it takes on all integer values in the other Coulomb phase. 
Based on this, we predict that experimental realizations with $|\mu| \ll J$ and negative will go from a $Z_2$ flux deconfined Coulomb phase to a $Z_2$ flux-confined Coulomb phase when they are cooled below this confinement transition, while realizations with $|\mu| \ll J$ and positive  will show signatures of a continuous $XY$-like transition from a flux-deconfined Coulomb liquid to a short-range correlated paramagnet.  

We establish these results via explict computations in a minimal model that captures the essential physics:
\begin{eqnarray}
H &=& \sum_{\langle r r' \rangle} (J_zS^z_r S^z_{r'} + J_{\perp}S^{\perp}_r \cdot S^{\perp}_{r'}) +\Delta \sum_r (S^z_r)^2 + \dots \; ,\nonumber \\
&&
\label{Hbasic}
\end{eqnarray}
where $z$ refers as before to the local easy-axis along the body diagonals of the pyrochlore tetrahedra (Fig.~\ref{fig:SchematicTestCorrelatorSpecificHeat}(a)), $\langle r r'\rangle$ denotes nearest-neighbor bonds of the pyrochlore lattice, the Ising exchange $J_z$, the easy-plane anisotropy $\Delta$, and their difference  $J_{\perp}$ $|J_z -\Delta|$ are all much larger than the transverse exchange interaction $J_{\perp}$, and the ellipses denote subdominant interactions of order $J_{\perp}$ or smaller.
With this in mind, we write $J_z = J$ and set $\Delta = J+\mu$, with the proviso that $|J_{\perp}| \ll |\mu| \ll J$. Below, we demonstrate that this model has the three phases advertised above, with a $Z_2$ flux confinement transition at $T_{c_2} \approx 1.57|\mu|$ (for negative $\mu$) and an $XY$-like transition at $T_{c_1} \approx 0.98 \mu$ (for positive $\mu$).



\section{Analysis}
At temperatures $|J_{\perp}| \ll T \ll J$, physical properties are controlled by thermal fluctuations {\em within the minimally-frustrated subset of configurations} that minimize the ${\mathcal O}(J)$ easy-axis exchange energy. This classical physics is controlled by the value of $\mu/T$. While the sign of this ratio is of course fixed for any given system, its magnitude can take on arbitrarily large values as the temperature is lowered below the scale set by $|\mu|$.  With this motivation, we focus here on the $T/J \rightarrow 0$ limit with $J_{\perp} =0$, and provide a precise analysis of the phase diagram in this limit as a function of the fugacity variable $w = \exp(-\mu/T)$. Of course, in a given system with a particular sign of $\mu$, one only has access to roughly one half of this phase diagram. Nevertheless, as we see below, this includes some interesting low-temperature physics.

All spin configurations $S^z_{r}$ that minimize the easy-axis exchange energy are characterized by the constraint that $S^z_{\rm tet}  = 0$ for each pyrochlore tetrahedron, where $S^z_{\rm tet} \equiv \sum_{r \in {\rm tet}} S^z_r$ is the total spin of a tetrahedron.
It is convenient to represent these in terms of a polarization field $\vec{P} = S^z_r \hat{d}_{AB}$ on oriented links $\hat{d}_{AB}$ of the diamond lattice whose $A$-sublattice ($B$-sublattice) vertices are the centers of up (down) pointing pyrochlore tetrahedra (Fig.~\ref{fig:SchematicTestCorrelatorSpecificHeat}(a)). In this language, the energetic constraint $S^z_{\rm tet} = 0$ is equivalent to $\vec{P}$ being divergence-free: $\Delta \cdot \vec{P} = 0$. There are three distinct ways of satisfying this constraint at the level of a tetrahedron: First, all four spins on a tetrahedron can have $S^z_r = 0$. Second, two of the spins can have $S^z_r=0$ while the other two take on values $+1$ and $-1$ respectively. Third, two of the spins can have $S^z_r = +1$ and the other two can have the opposite sign. Using the fact that each pyrochlore spin is shared between two tetrahedra, it is easy to see that the Boltzmann weight of a configuration acquires a factor of $w$ (factor of $1$) for each spin $S^z_r= \pm 1$ ($S^z_r =0$). This low-temperature physics is encoded in the partition function 
\begin{equation}
Z = \sum_{C} w^{n_{1}(C)} \; ,
\end{equation}
where the sum is over all minimally-frustrated configurations $C$ that satisfy zero-divergence constraint on $\vec{P}$ and $n_{1}(C)$ is the number of pyrochlore spins in a configuration $C$ that have $S^z_r = \pm 1$.
\begin{figure*}
    (a)\includegraphics[width=0.3\linewidth]{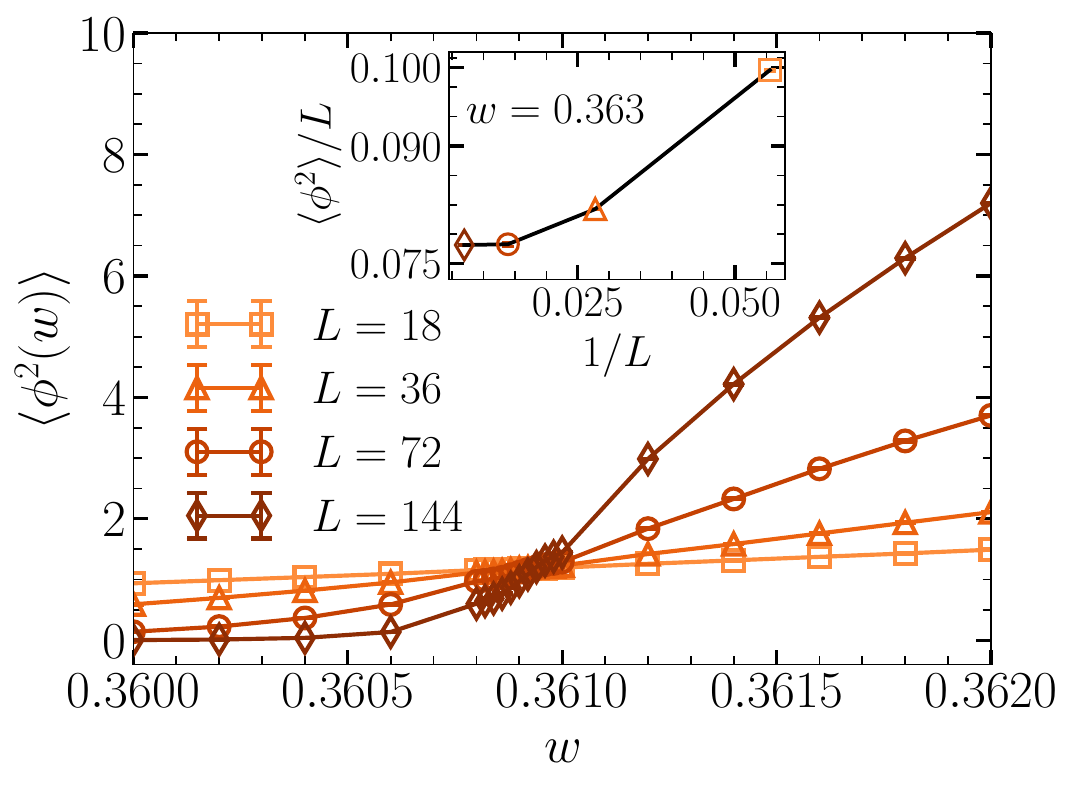}
    (b)\includegraphics[width=0.3\linewidth]{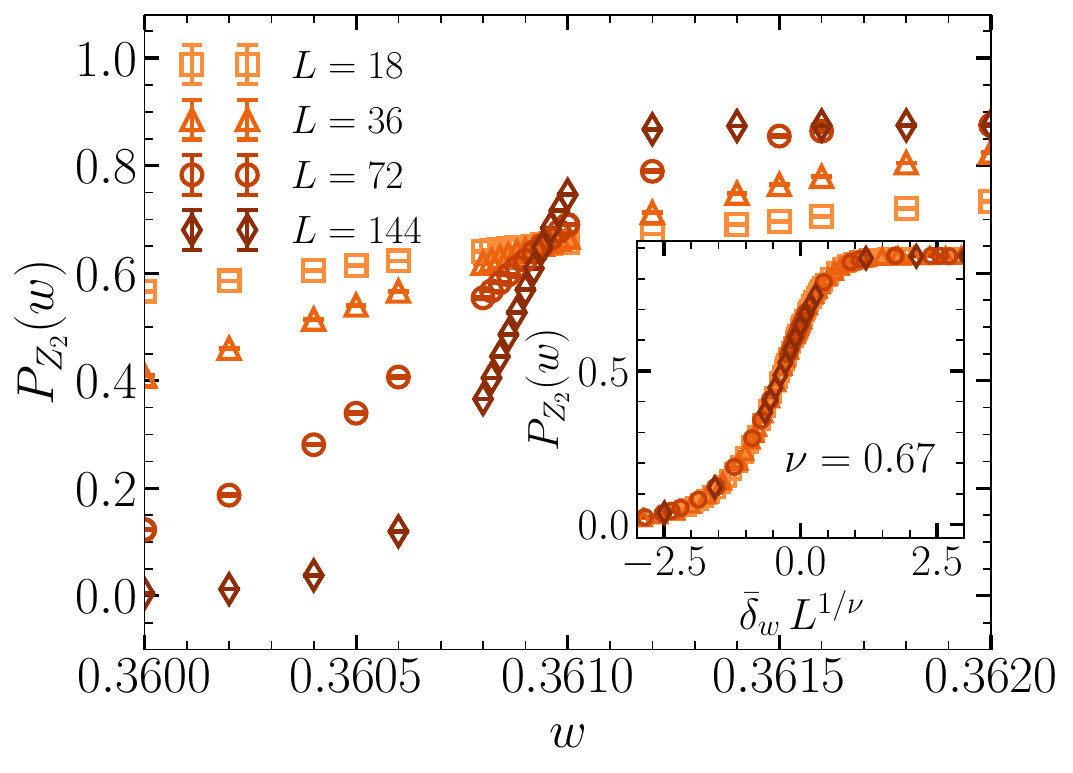}
    (c)\includegraphics[width=0.3\linewidth]{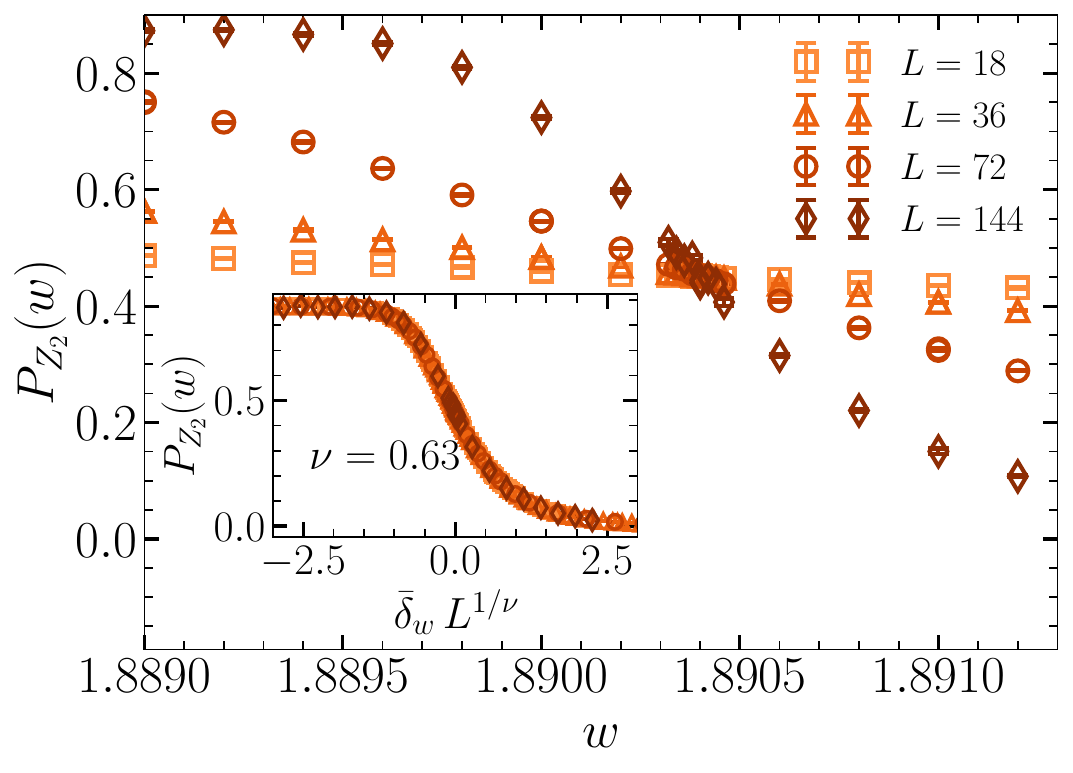}
    \caption{(a) $\langle \phi^2 \rangle (w) $ shows a sharp crossing at $w_{c_1} =0.3609(1)$ of curves corresponding to different $L$, with $\langle \phi^2 \rangle \sim L$ for $w> w_{c_1}$ (inset), and $\langle \phi^2 \rangle \rightarrow 0$ for $w< w_{c_1}$.  (b) and (c) $P_{Z_2}(w)$, the probability that the $Z_2$ flux is nonzero, shows two such crossings, one at $w_{c_1} =0.3609(1)$ and the other at $w_{c_2} = 1.8904(1)$. When plotted as a function of $\delta w L^{1/\nu}$  with
    $\delta w \equiv w-w_{c1}$ ($\delta w \equiv w -w_{c_2}$) and $\nu \approx0.67 (3)$ ($\nu \approx0.63(3) $) in the vicinity of these crossings,  data for $P_{Z_2}$ exhibits a scaling collapse}
    \label{fig:phisqrPfrac}
\end{figure*}

Clearly, $w=0$ is
 special, since the only configurations that contribute have $S^z_r =0$ and the system is
simply non-magnetic. However, since there is an entropic advantage to having nonzero values of $S^z_r$ at any nonzero $w$, one might expect that this behavior does not persist. In the other limit of large $w \gg 1$, one certainly expects (by analogy to classical $S_{\rm eff} =1/2$ spin-ice) Coulomb spin liquid behavior described by fluctuations of a coarse-grained divergence-free polarization field. Further, since it seems reasonable that this long-wavelength physics does not really depend on microscopic details such as the allowed values for the lattice-level polarization field, one might expect that this large-$w$ Coulomb spin liquid phase would extend to nonzero $w \ll 1$. 

Surprisingly, our results below demonstrate that this is not the case:
Not only is the large-$w$ regime {\em not} continuously connected to the small-$w$ regime, it turns out that the phase diagram has not one, but {\em two} different thermodynamic phase transitions that give rise to three distinct equilibrium phases as a function of $w$.

\section{Results}
We use Monte-Carlo simulations of $L \times L \times L$ samples (comprising $4L^3$ sites) with periodic boundary conditions imposed along the principal axes (Fig.~\ref{fig:SchematicTestCorrelatorSpecificHeat}(a)) to map out the phase diagram; these rely, mutatis mutandis, on the myopic~\cite{Rakala_Damle_2017,Morita_Lee_Damle_Kawashima2023,Kundu_Damle_PRX_2025,Kundu_Damle3Dpreprint2025} worm algorithms~\cite{Alet_Sorenson2003,Sandvik_Moessner_2006} developed and documented in earlier work. 
The basic idea is to create a pair of oppositely-charged violations of the $S^z_{\rm tet} =0$ constraints on next-nearest neighbor tetrahedra, and construct a biased random walk for one of them (the worm head) keeping the other (worm tail) fixed. The walk ends when the wandering head reunites with the fixed tail, returning the system to a different minimally-frustrated configuration. The bias satisfies detailed balance, ensuring that each update step can be accepted with unit probability upon completion of the walk. In our problem, one can do this with with a pair of $\Delta S^z_{\rm tet} = \pm 1$ defects, which are the elementary charge excitations in our problem, or, in a more restricted way with  $\Delta S^z_{\rm tet} = \pm 2$ defects. A key advantage of this algorithm is that the histogram $h^{(1)}(\vec{r})$ of the head-to-tail displacements for the unit-charge worm corresponds exactly to the equilibrium correlation function of a pair of $\pm 1$ test charges~\cite{Kundu_Damle_PRX_2025,Kundu_Damle3Dpreprint2025,Alet_Sorenson2003}. Although the analogous histogram $h^{(2)}(\vec{r})$ for charge-$2$ worms is not the estimator for the corresponding test charge correlator, deconfined  behavior of $h^{(2)}(\vec{r})$ does imply the same for the test charge correlator~\cite{Kundu_Damle_PRX_2025,Kundu_Damle3Dpreprint2025}.
\begin{figure*}
    (a)\includegraphics[width=0.22\linewidth]{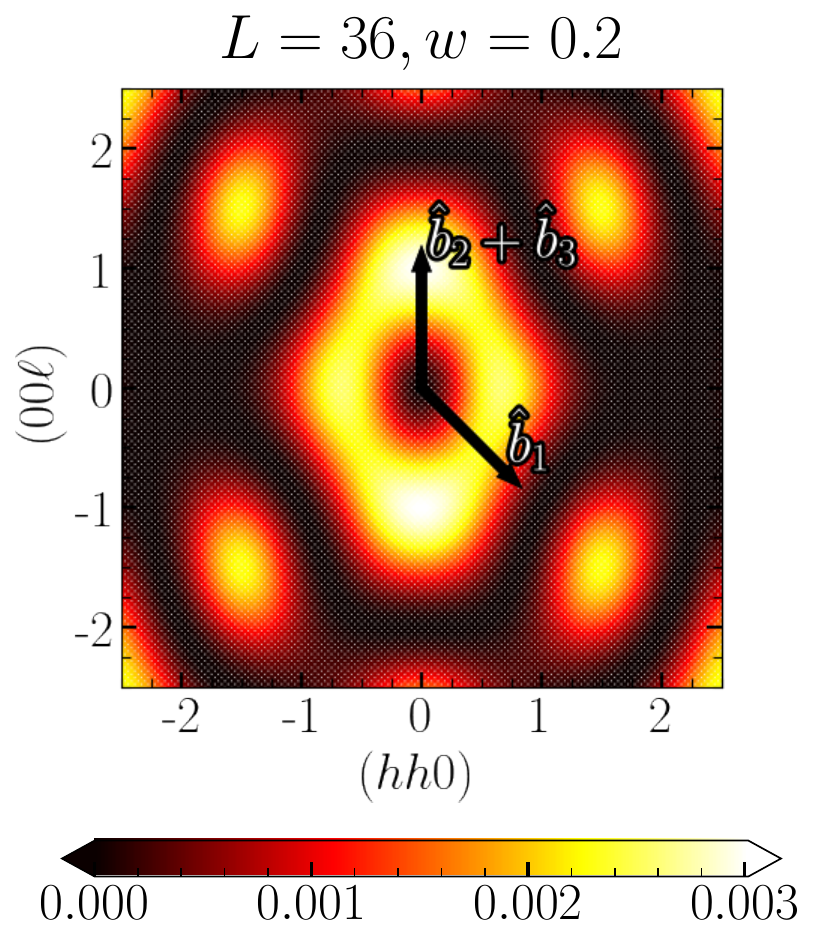}
    (b)\includegraphics[width=0.22\linewidth]{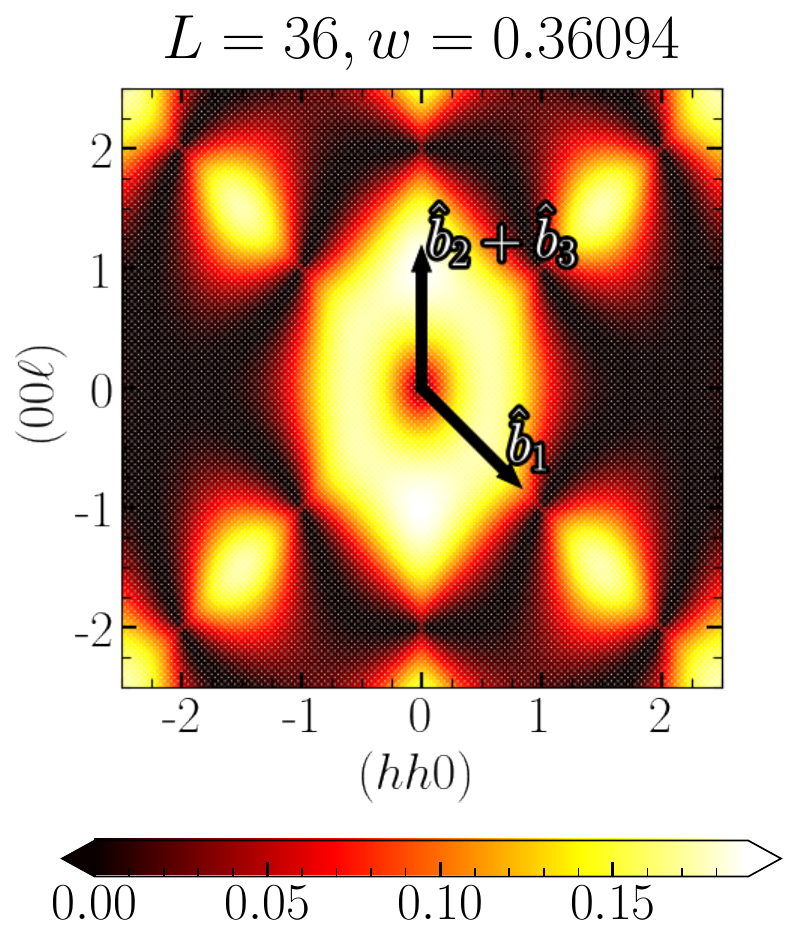}
    (c)\includegraphics[width=0.22\linewidth]{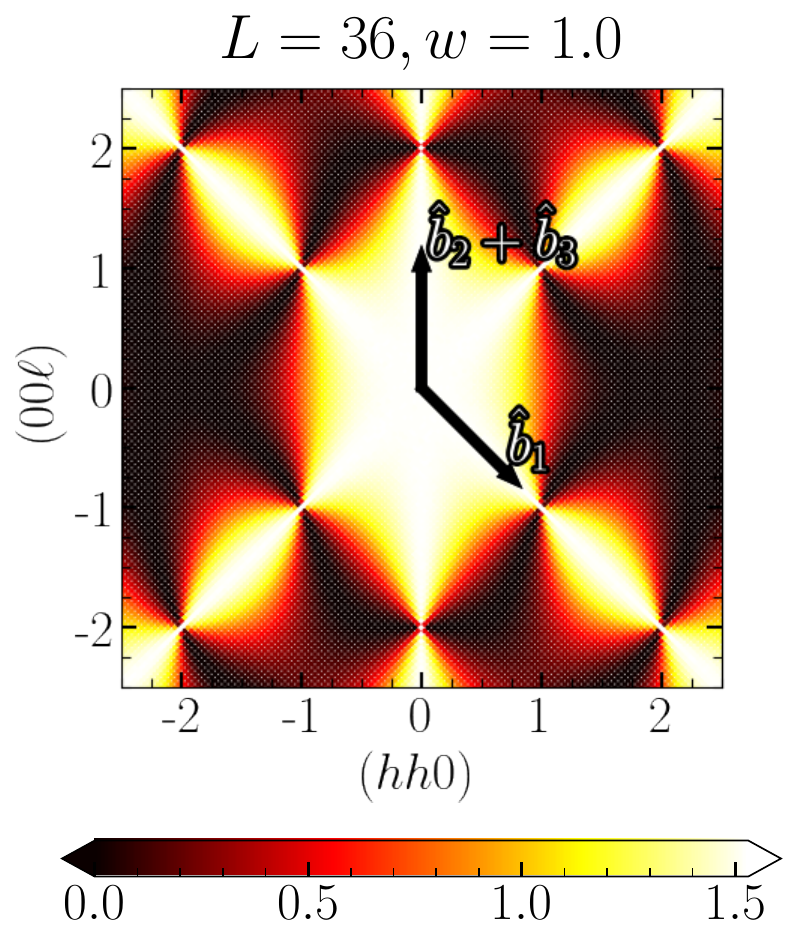}
    (d)\includegraphics[width=0.22\linewidth]{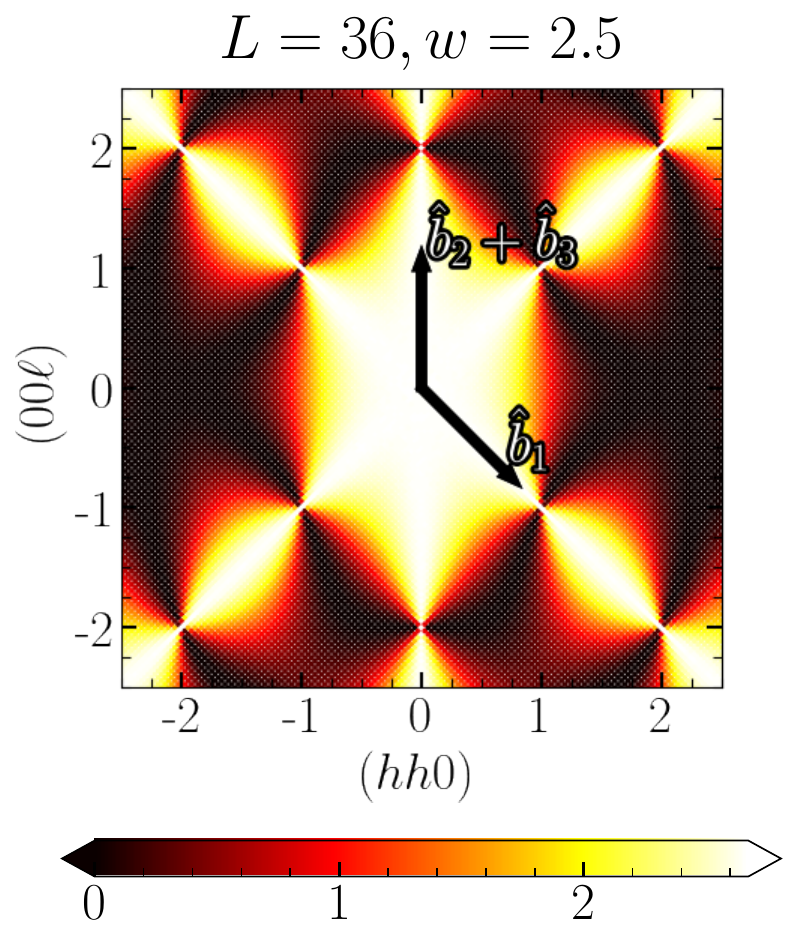}
    \caption{(a) For $w<w_{c_1}$, the spin flip structure factor~\cite{Chung_etal2010} $S(\vec{q})$ (with $\vec{q} = [h,h,l]$ in cubic coordinates) is featureless, as expected for a non-magnetic or paramagnetic state. (b) At $w_{c_1}$, it shows the first hints of incipient dipolar structure. (c), (d) For all $w>w_{c_1}$, it shows clear signatures of a Coulomb liquid phase, with its characteristic pinch-point singularities. $\hat{b}_i$ are reciprocal lattice directions dual to lattice translations $\vec{a}_{i}$ (Fig.~\ref{fig:SchematicTestCorrelatorSpecificHeat}).}
    \label{fig:StructFact}
\end{figure*}

The first clue that there are three distinct phases comes from the fact that $h^{(1)} (\vec{r})\rightarrow 0$ for large $r$ both at small $w \ll 1$ and at very large $w \gg 1$, but it goes to a nonzero constant at large $r$ for intermediate $w \sim 1$ (Fig.~\ref{fig:SchematicTestCorrelatorSpecificHeat}(b)). 
From  Fig.~\ref{fig:SchematicTestCorrelatorSpecificHeat}(c),
we also see that the specific heat $c_v =  \log^2(w)\langle \delta n_{1})^2 \rangle/L^3$, where $\langle \delta n_{1})^2 \rangle$ is the mean square fluctuation of the number $n_1$ of $S^z=\pm1$ spins, also carries signatures of two distinct thermodynamic transitions.

To characterize these distinct phases, we study $\vec{\phi} \equiv (\phi_1,\phi_2,\phi_3)$, the fluxes of $\vec{P}$ along the principal axes of our periodic sample (Fig.~\ref{fig:SchematicTestCorrelatorSpecificHeat}(a)). These are integer-valued and unaffected by local changes to a configuration, as is clear from the following: A configuration with $\vec{\phi} = 0$ can only change its flux vector if a worm head winds around a periodic direction before annihilating with the tail; this necessarily leads to a {\em global} change of the spin configuration. Further, since the smallest such change involves a charge $\Delta S^z_{\rm tet} = \pm 1$ worm, Gauss's law implies that only integer-valued fluxes are possible.  

Next we note that  components of $\vec{\phi}$ {\em cannot change by $\pm 1$} in the thermodynamic limit in any phase in which $h^{(1)}(\vec{r})$ decays rapidly to zero at large $r$ (as is the case for $w \gg 1$ and $w \ll 1$).
In the thermodynamic limit, there are two possible scenarios for such a phase: Either $\vec{\phi}$ could simply be pinned to $\vec{\phi} = 0$, or it could fluctuate with components restricted to be even integers. 
Since $\vec{\phi} = 0$ identically at $w=0$, the first scenario is plausible for small $w \ll 1$. Similarly, the second scenario is more likely at large $w \gg 1$ since $h^{(2)}(\vec{r})$ is deconfined in this limit (End-Matter).
To examine the first possibility, we monitor the expectation value $\langle \phi^2 \rangle$, where
\begin{eqnarray}
\phi^2 & \equiv & (\phi_1^2 +\phi_2^2 +\phi_3^2 +\phi_4^2)/2 \; ,
\end{eqnarray}
and $\phi_4 = -(\phi_1 +\phi_2 +\phi_3)$. In addition, motivated by the second possibility, we keep track of $P_{\rm Z_2}$, the probability that at least one of the components of $\vec{\phi}$ is an {\em odd} integer.

When $\langle \phi^2 \rangle$
is plotted at fixed $L$ as a function of $w$, curves corresponding to different $L$ all
cross at $w_{c_1} \approx 0.3609(1)$: 
For $w < w_{c_1}$, $\langle \phi^2 \rangle \rightarrow 0$ with increasing $L$, while
$\langle \phi^2 \rangle \sim L$ at large $L$ for  $w > w_{c_1}$ (Fig.~\ref{fig:phisqrPfrac}(a) and inset). The corresponding data for
$P_{Z_2}$ shows two such sharp crossings, one at $w_{c_1}$, and another at $w_{c_2} \approx1.8904 (1)$:
For all $w \in [w_{c_1},w_{c_2}]$, $P_{Z_2}$ goes to a
 nonzero limit with increasing $L$, but decays to zero at large $L$ for all $w \notin [w_{c_1},w_{c_2}]$  (Fig.~\ref{fig:phisqrPfrac}(b) \ref{fig:phisqrPfrac}(c)). When plotted as a function of $\bar{\delta}_w L^{1/\nu}$ in the vicinity of $w_{c_1}$ ($w_{c_2}$), where $\bar{\delta}_w = (w-w_{c_1})/w_{c_1}$  and $\nu \approx 0.67(3)$ ($\bar{\delta}_w = (w-w_{c_2})/w_{c_2}$ and $\nu \approx 0.63(3)$), our data collapses on to a single scaling curve (Fig.~\ref{fig:phisqrPfrac}(b),  \ref{fig:phisqrPfrac}(c) insets). Since this is the expected behavior for a dimensionless quantity like $P_{Z_2}$ in the vicinity of a continuous phase transition~\cite{Cardy_1996}, we conclude that there are indeed three distinct phases, separated by critical points at $w_{c_1}$ and $w_{c_2}$ 

The divergence-free polarization field $\vec{P}$ is analogous to the divergence-free current in the 
link-current representation of a classical three-dimensional 
(3D) $XY$ model (equivalently, a two-dimensional bosonic system at $T=0$)~\cite{Alet_Sorenson2003}. In this analogy, $\langle \phi^2\rangle /L$ is exactly analogous to the Monte Carlo estimator for the spin stiffness $\rho_s$ of the 
$XY$ model, and the unit test charge correlator obtained form $h^{(1)}(\vec{r})$ maps to the spin correlation $\langle e^{i\theta(\vec{r})} e^{-i\theta(0)} \rangle$ of fictitious $XY$ spins $e^{i\theta(\vec{r})}$~\cite{Alet_Sorenson2003}.  

Thus, the sharp crossing at $w_{c_1}$ is exactly what one would expect from this viewpoint, since standard scaling ideas predict that $L\rho_s$ is dimensionless at the 3D-$XY$ critical point~\cite{Cardy_1996}.
Consistent with our earlier estimate of $\nu \approx 0.67(3)$, this argument predicts that the transition at $w_{c_1}$ corresponds to the ferromagnetic ordering transition of the 3D $XY$ model. It also implies that the spin structure factor at $w_{c_1}$ provides an unusual probe of critical current correlations at an $XY$-like transition.
  
This transition is clearly visible in the spin correlations: We find that the low-$w$ phase is nonmagnetic with a featureless spin flip structure factor, readily distinguishable from the critical behavior seen at $w_{c_1}$ (Fig.~\ref{fig:StructFact}(a), \ref{fig:StructFact}(b))). For all $w> w_{c_1}$, the spin flip structure factor exhibits dipolar pinch-point singularities characteristic of a Coulomb spin liquid, with no obvious signature of the transition at $w_{c_2}$
  (Fig.~\ref{fig:StructFact}(c),(d)). 
  
  Thus, both phases on either side of $w_{c_2}$ are Coulomb spin liquids. This is
   further confirmed by a detailed study of the probability distribution of the flux vector 
   $\vec{\phi}$: This is a Gaussian for all $w>w_{c_1}$, consistent with expectations for a 
   Coulomb liquid. For all $w \in (w_{c_1}, w_{c_2})$, all integer vectors $\vec{\phi}$ obey this Gaussian behavior, whereas only $\vec{\phi}$ with even-integer components lie on this Gaussian 
   curve for $w>w_{c_2}$ and any $\vec{\phi}$ with an odd-integer component has vanishing probability in the thermodynamic limit (see Appendix).
The transition at $w_{c_2}$ thus involves a confinement of the $Z_2$ part of the integer-valued flux of the Coulomb field. 

The full machinery of a $Z_2$ gauge theory is not needed to understand this $Z_2$ flux confinement transition: To see why, recall that the unit test charge correlator obtained from $h^{(1)}(\vec{r})$ is confined for $w>w_{c_2}$, but the deconfined behavior of $h^{(2)}(\vec{r})$ implies that the charge-$2$ test correlator remains deconfined for $w>w_{c_2}$. In the $XY$ model, the latter maps to the correlation function $\langle e^{2i\theta(\vec{r})} e^{-2i\theta(0)} \rangle$~\cite{Alet_Sorenson2003}, which is thus long-range ordered for $w>w_{c_2}$, although $\langle e^{i\theta(\vec{r})} e^{-i\theta(0)} \rangle$ is short-ranged. Thus, the transition at $w_{c_2}$ must be in the universality class of the ferromagnet-nematic transition of the 3D $XY$ model~\cite{Cardy_1996}, {\em i.e.} an Ising transition. This fits in with the $Z_2$ flux-confinement interpretation since the Ising gauge theory and spin models are dual to each other in three dimensions. It is also consistent with our estimate of the correlation length exponent $\nu \approx 0.63(3)$ for the transition at $w_{c_2}$. The Appendix provides additional evidence in support of these idenfications.

\section{Discussion}
This {\em competition} between easy-axis exchange anisotropy and easy-plane single-ion anisotropy is also at the heart of a previous proposal~\cite{Kundu_Damle_PRX_2025} for unusual low-temperature behavior on the one-third magnetization plateau in spin $S=1$ kagome magnets, but accessing it would require a strong external field $B \sim J$ perpendicular to the kagome plane. In contrast, the pyrochlore physics that we focus on here arises in zero magnetic field, making it potentially more accessible to experiments. We thus hope our work motivates attempts to identify materials with such competing anisotropies. Finally, we note that another motivation for identifying the ingredients that lead to such competition comes from parallel work~\cite{Pandey_Kundu_Damle_3by2preprint2025} that makes the case for two topologically-distinct Coulomb phases separated by a low-temperature $Z_3$ flux confinement transition in effective $S=3/2$ pyrochlore magnets with such competing anisotropies.

\section*{Acknowledgements}
We thank Subhro Bhattacharjee, Radu Coldea and Souvik Kundu for useful discussions and Souvik Kundu for collaboration on closely related work~\cite{Pandey_Kundu_Damle_3by2preprint2025}. We gratefully acknowledge generous allocation of computing resources by the Department of Theoretical Physics (DTP) of the Tata Institute of Fundamental Research (TIFR), and related technical assistance from K. Ghadiali and A. Salve. The work of JP was supported at the TIFR by a graduate fellowship from DAE, India.  KD was supported at the TIFR by DAE, India, and in part by a J.C. Bose Fellowship (JCB/2020/000047) of SERB, DST India, and by
	the Infosys-Chandrasekharan Random Geometry Center
	(TIFR).

\bibliography{bibliography}

\section*{Appendix}

\begin{figure*}
    (a)\includegraphics[width=0.22\linewidth]{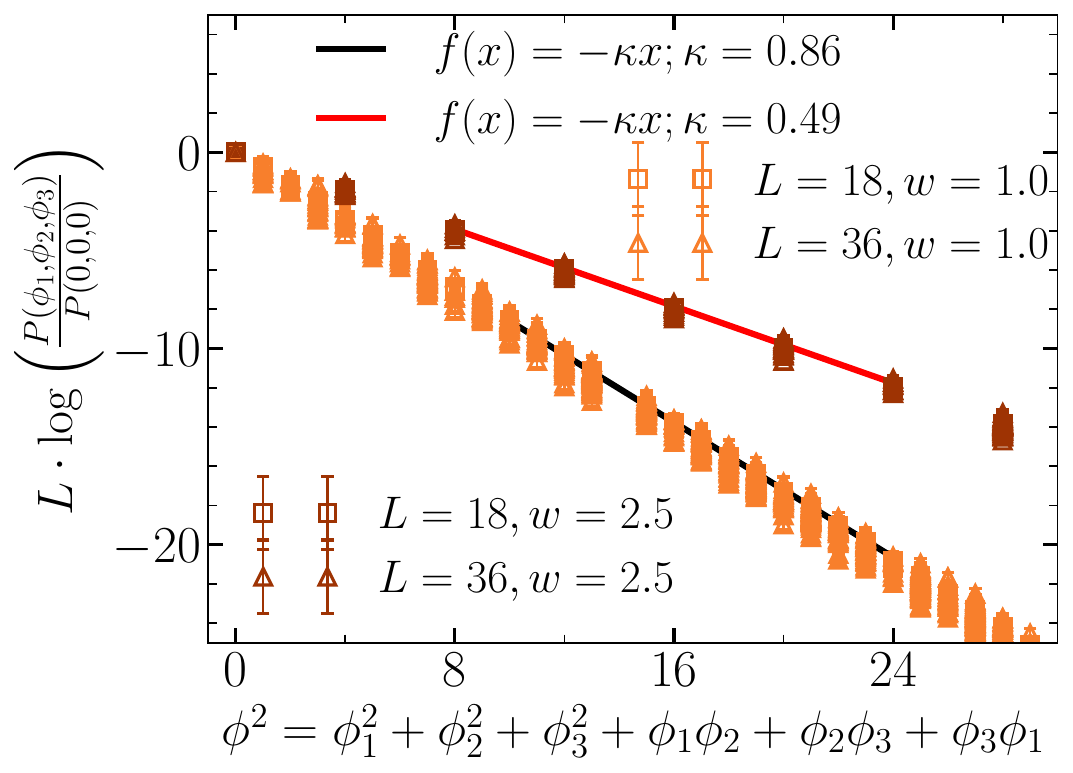}
    (b)\includegraphics[width=0.22\linewidth]{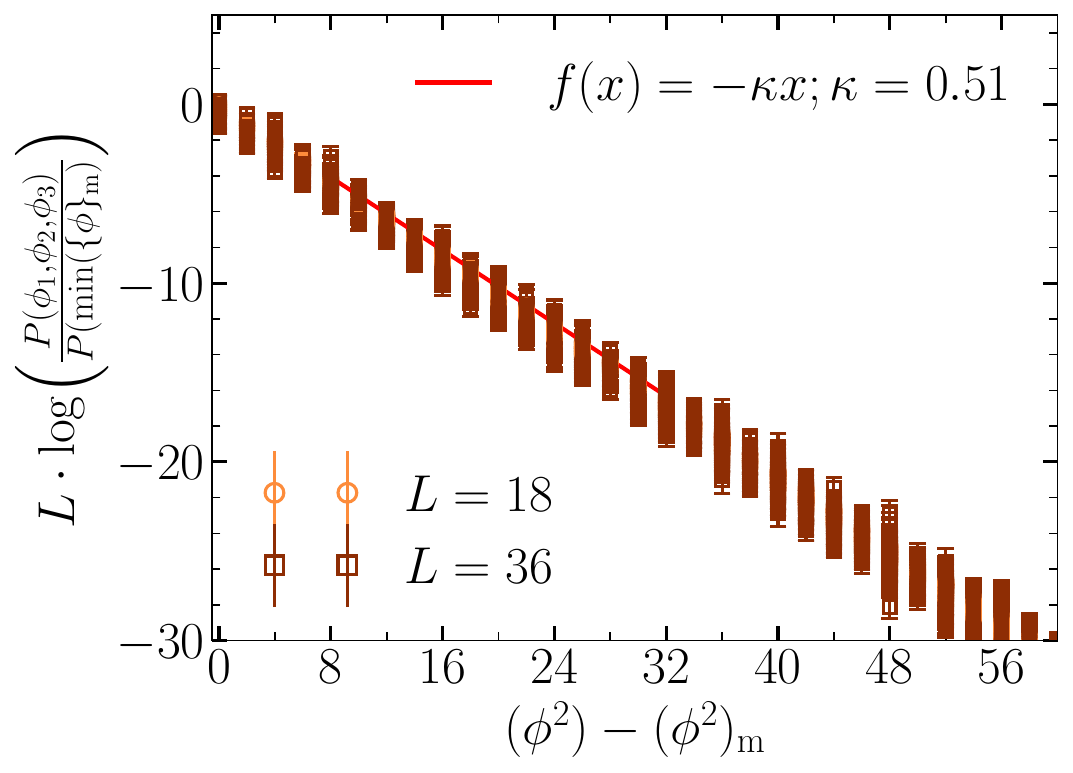}
    (c)\includegraphics[width=0.22\linewidth]{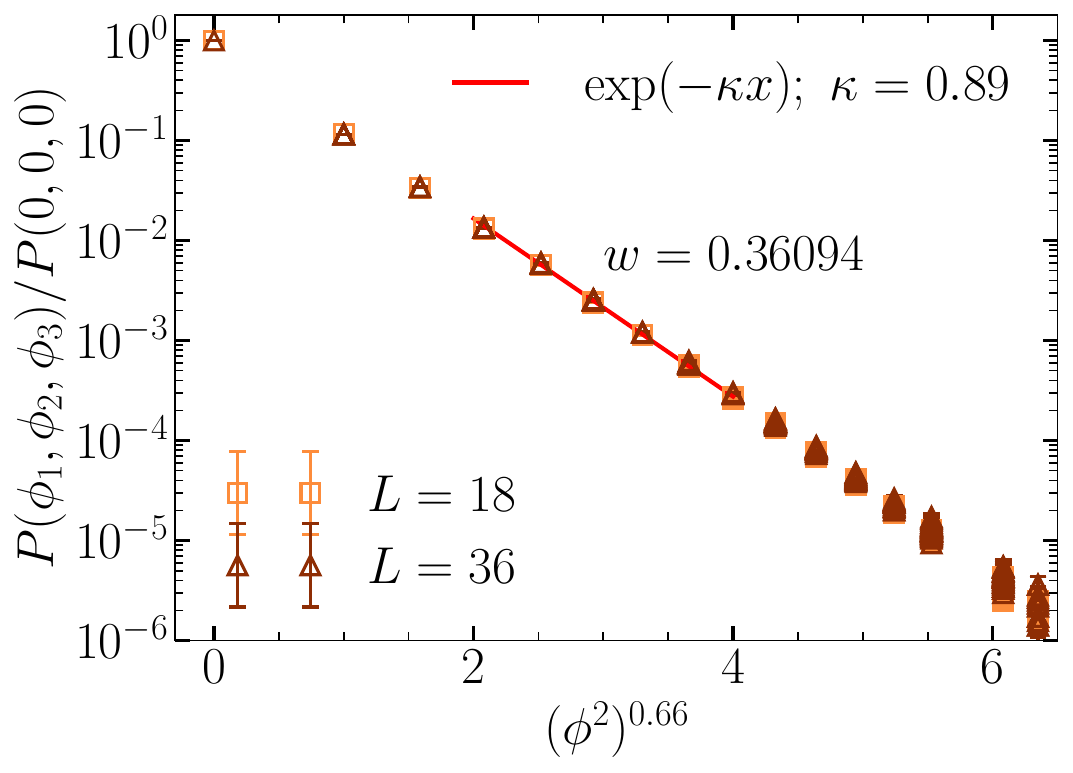}
    (d)\includegraphics[width=0.22\linewidth]{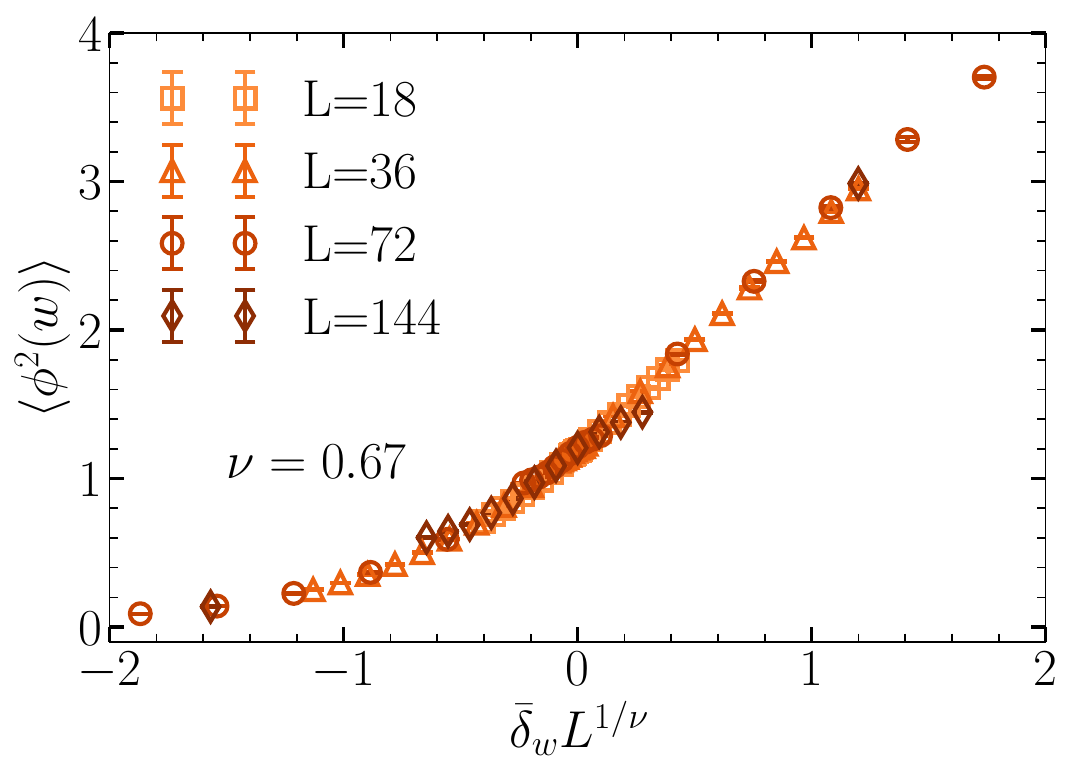}
    \caption{(a) The flux distribution $P(\vec{\phi})$ is a Gaussian, $P(\vec{\phi}) \propto \exp(-\kappa \phi^2)$,  for all $w>w_{c_2}$ with the exception of $w=w_{c_2}$, with $\phi^2$ defined as in the main text. For $w>w_{c_2}$, only      $\vec{\phi}$ with even integer components obeys this Gaussian distribution and all other flux vectors are
      suppressed in the thermodynamic limit, while all integer flux vectors obey the Gaussian when 
      $w \in (w_{c_1}, w_{c_2})$. (b) At $w_{c_2}$, fluxes form sectors $\{\phi\}_m$ ($m=0,2,3$) distinguished by
       the number $m$ of components of $\vec{\phi}$ that are odd integers. $\vec{\phi}$ belonging to different sectors obey distinct Gaussian distributions with a common stiffness $\kappa/L$ but different
         prefactors. To test this, one plots 
     $L\log\left( \frac{ P(\vec{\phi})}{P({\rm min}\{ \phi \}_m)} \right)$ versus $\phi^2 - \phi^2_m$, where ${\rm min} \{ \phi \}_m$ denotes a flux configuration in sector $m$ with  $\phi^2$ equal to $\phi^2_m$, the minimum possible value of $\phi^2$ in sector $m$. (c) At $w_{c_1}$, we find that the distribution $P(\vec{\phi})$ can be approximately modeled by a stretched exponential function of $\phi^2$. (d) In the vicinity of $w_{c_1}$, $\langle \phi^2 \rangle$ for various $L$ obeys a finite-size scaling form $\langle \phi^2 \rangle = f(\bar{\delta}_w L^{1/\nu})$, where $\bar{\delta}_w \equiv (w-w_{c_1})/w_{c_1}$ and $\nu \approx 0.67(3)$. 
    }
    \label{fig:EndMatFluxes}
\end{figure*}
As mentioned in the main text, the probability distribution $P(\vec{\phi})$ of the flux vector $\vec{\phi}$ is a Gaussian for all $w> w_{c_1}$, with width scaling as $L$. 
However, when $w>w_{c_2}$, only $\vec{\phi}$ with all components being even integers obey this Gaussian distribution in the thermodynamic limit, while all integer-valued flux vectors obey this Gaussian distribution when
 $w \in (w_{c_1},w_{c_2})$.
This is illustrated by the data shown in Fig.~\ref{fig:EndMatFluxes}(a).

Precisely {\em at} $w_{c_2}$, this distribution has a much more intricate form, entirely analogous to that explored in Ref.~\cite{Kundu_Damle3Dpreprint2025}: The flux vectors $\vec{\phi}$ fall into five distinct sectors, distinguished by the number $m$ of components of $\vec{\phi}$ that are odd integers, where the distinct choices for $m$ are $m=0,2,3$ ($m=1$ is equivalent to $m=2$ due to the symmetries of the diamond lattice). Each sector has its own Gaussian distribution of $\vec{\phi}$, with the stiffness in all sectors being $\kappa/L$ with a common value of $\kappa$ for samples of linear size $L$. The sectors are thus only distinguished by the prefactor of these Gaussians in each sector, which fixes their relative sectorial weight. This is shown in Fig.~\ref{fig:EndMatFluxes}(b). 

In contrast, at $w_{c_1}$, it appears to follow a stretched exponential form, expected to be characteristic of the current variable at a three-dimensional $XY$ transition by the arguments given in the main text. This is shown in Fig.~\ref{fig:EndMatFluxes}(c). Also, we find that $\langle \phi^2 \rangle$ in the vicinity of $w_{c_1}$ obeys the finite-size scaling form expected for $L\rho_s$ at the three-dimensional $XY$ transition (where $\rho_s$ is the spin stiffness). This is shown in Fig.~\ref{fig:EndMatFluxes}(d). Again, this is consistent with the arguments presented in the main text.

\begin{figure*}
   (a)\includegraphics[width=0.22\linewidth]{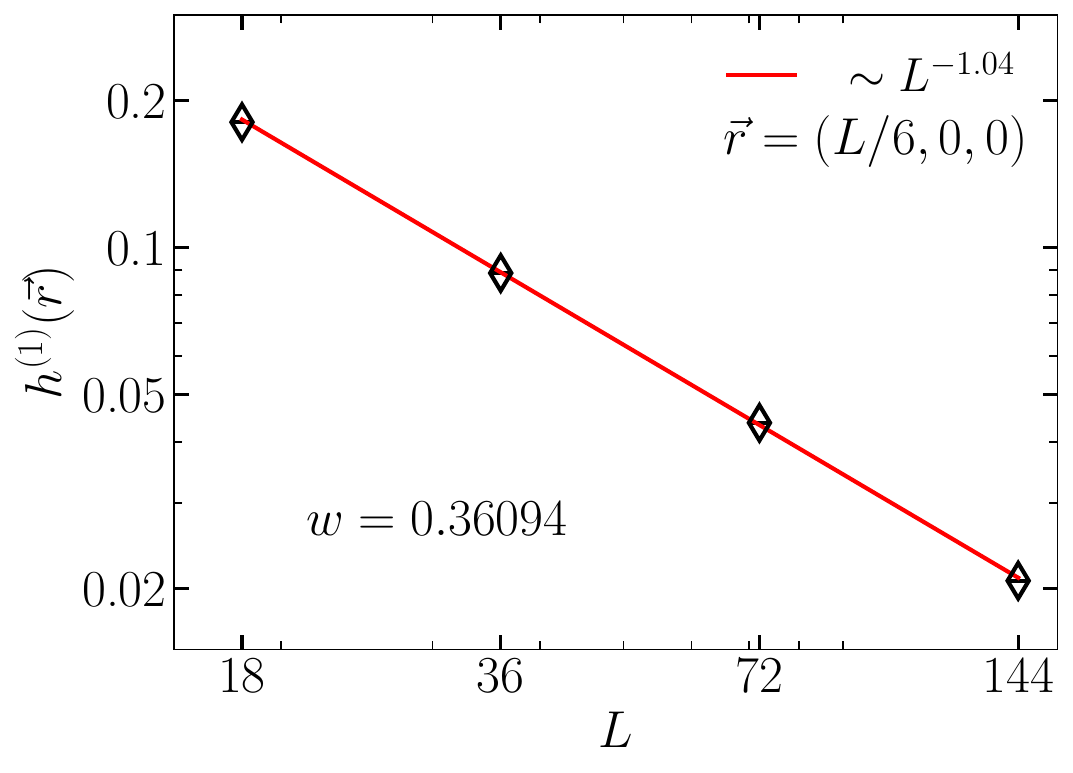}
    (b)\includegraphics[width=0.22\linewidth]{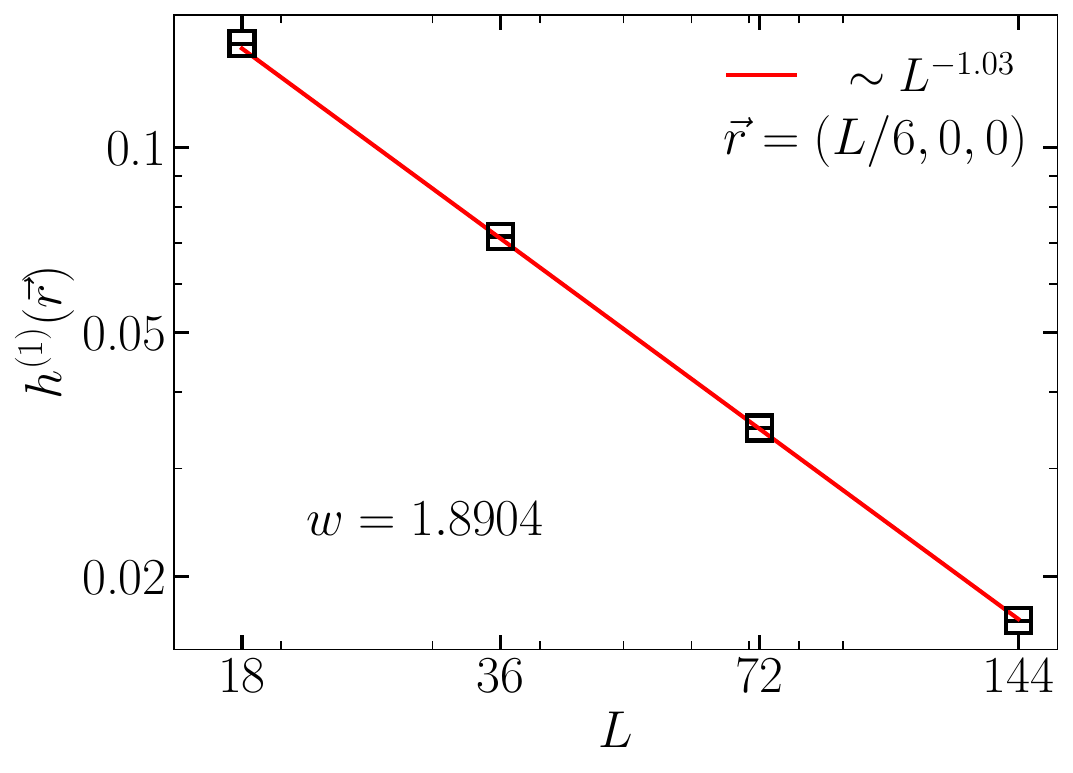}
    (c)\includegraphics[width=0.22\linewidth]{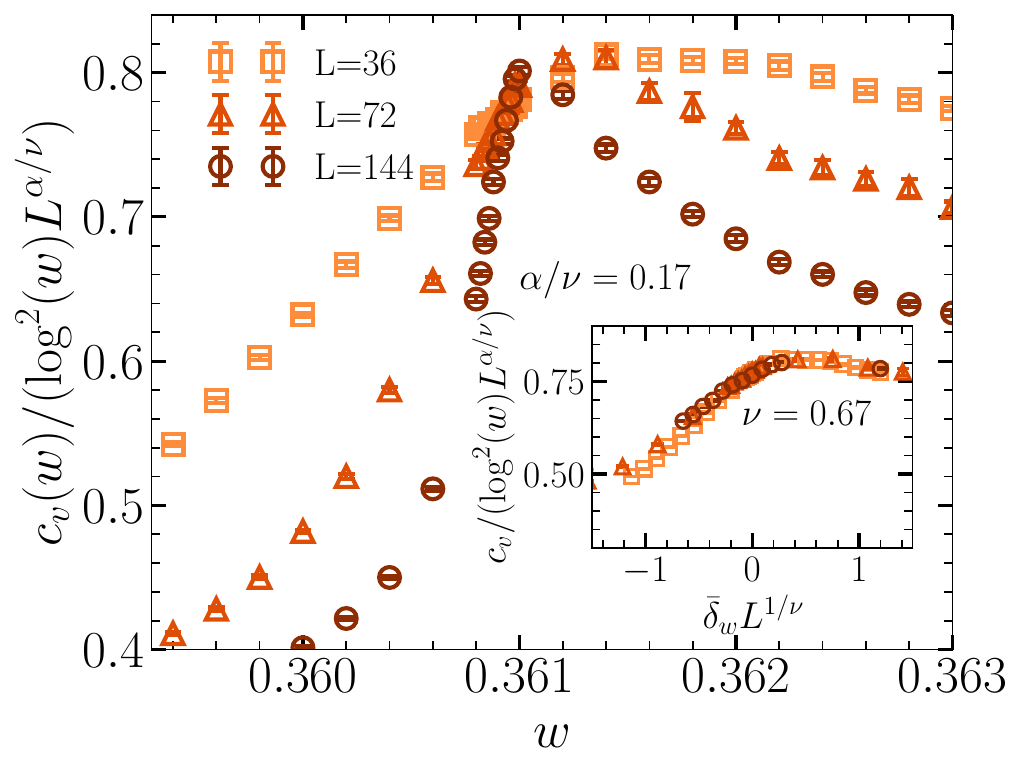}
    (d)\includegraphics[width=0.22\linewidth]{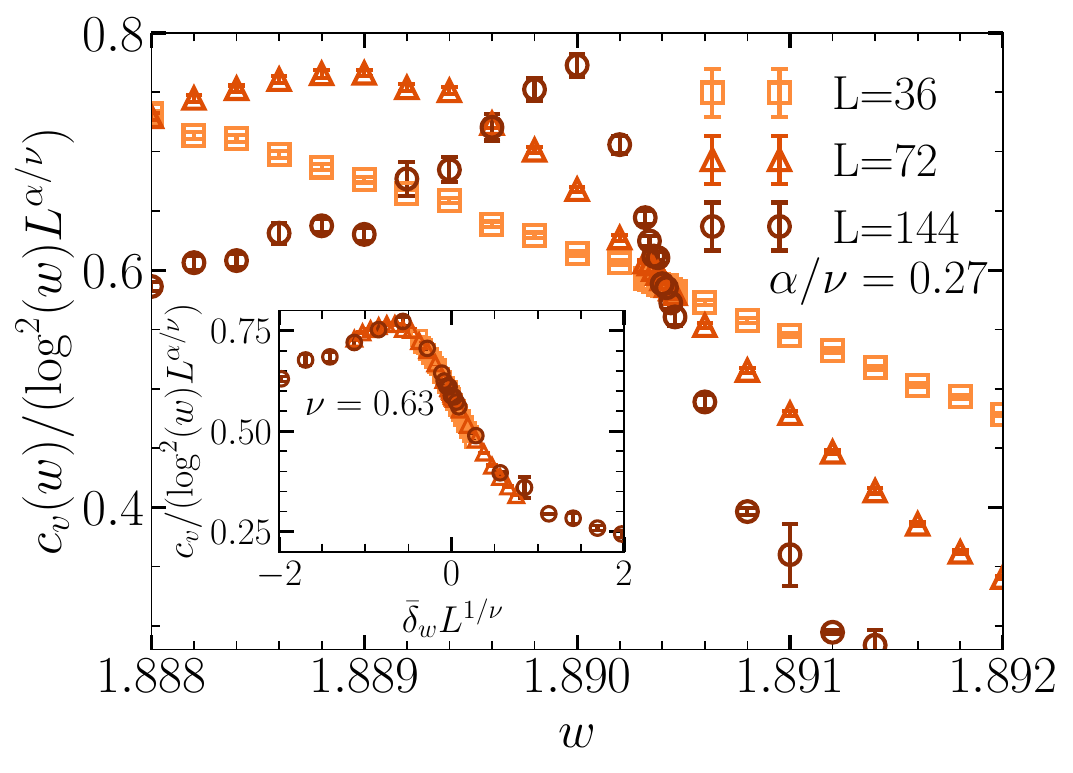}
    \caption{(a), (b) The histogram $h^{(1)}(\vec{r})$ of head-to-tail displacements of unit-charge worms, which serves as the estimator for the unit test charge correlator $C^{(1)}(\vec{r})$, has a power-law form $\sim 1/r^{1+\eta}$ at $w=w_{c_1}$ ($w=w_{c_2}$), with $\eta = 0.04(3)$ ($\eta =0.03(3) $). (c), (d) In the vicinity of $w_{c_1}$ ($w_{c_2}$), the specific heat $c_v$ for various $L$ obeys a finite-size scaling form $c_v = L^{\alpha/\nu} g(\bar{\delta}_w L^{1/\nu})$, where $\bar{\delta}_w \equiv (w-w_{c_1})/w_{c_1}$ ($\bar{\delta}_w \equiv (w-w_{c_2})/w_{c_2}$)and $\nu \approx $ 0.67(3)($\nu \approx 0.63(3)$).}
    \label{fig:TestChargeSpecificHeat}
\end{figure*}

We have also studied the $r$ dependence of the estimator $h^{(1)}(\vec{r})$ for the unit test charge correlator at $w_{c_1}$ and $w_{c_2}$, finding power-law behavior in both cases. 
Although our range of $L$ does not allow us to make a very precise determination of the anomalous exponents $\eta$ in each case, we see that our estimates for these power-law exponents at $w_{c_1}$ and $w_{c_2}$ are consistent with the expected behavior of the spin correlation function at the three-dimensional $XY$ and Ising critical points respectively. This is shown in Fig.~\ref{fig:TestChargeSpecificHeat}(a), \ref{fig:TestChargeSpecificHeat}(b).

Finally, we have studied the scaling of the specific heat in the vicinity of the transitions at $w_{c_1}$ and $w_{c_2}$. This is shown in Fig.~\ref{fig:TestChargeSpecificHeat}(c), \ref{fig:TestChargeSpecificHeat}(d). From this data analysis, we see that the the specific heat at various sizes $L$, when rescaled by $L^{\alpha/\nu}$ and plotted as a function of $w$ shows a clear crossing at $w_{c_1}$ ($w_{c_2}$) when the exponent is chosen to be $\alpha/\nu \approx0.17 (4)$ ($\alpha/\nu \approx 0.27(4)$). Since we have already estimated the correlation length exponent for the transition at $w_{c_1}$ ($w_{c_2}$) to be $\nu \approx 0.67(3)$ ($\nu \approx0.63(3) $), this is not consistent with hyperscaling, since hyperscaling demands that $\alpha = 2-3\nu$ in three dimensions.  

Thus, although the critical scaling behavior in the vicinity of the transitions was not our focus in the present work, we have found reasonably good evidence in favour of our theoretical arguments which place the transitions at $w_{c_1}$ and $w_{c_2}$ in  three-dimensional $XY$ and Ising universality classes respectively. As for the discrepancy seen in the scaling of the specific heat, we note that this could be the effect of a finite-size crossover. We hope to clarify this in future work.

\end{document}